\documentclass[aps, prd, superscriptaddress, nofootinbib, preprintnumbers]{revtex4}
\usepackage{amsmath}
\usepackage{graphicx}
\usepackage{color} 

\newcommand{\lesssim}{\mathrel{\mathpalette\vereq<}}
\newcommand{\gtrsim}{\mathrel{\mathpalette\vereq>}}

\newcommand{\chushi}[1]{}

\begin{document}
 \title{{\bf Dark Side of the Standard Model: Dormant New Physics Awaken}
 \vspace{5mm}}

\author{Shinya Matsuzaki}\thanks{\tt synya@hken.phys.nagoya-u.ac.jp}
      \affiliation{ Institute for Advanced Research, Nagoya University, Nagoya 464-8602, Japan.}
      \affiliation{ Department of Physics, Nagoya University, Nagoya 464-8602, Japan.}
\author{{Hiroshi Ohki}} \thanks{
      {\tt ohki@kmi.nagoya-u.ac.jp}}
      \affiliation{ RIKEN BNL Research Center, Brookhaven National Laboratory, Upton, NY 11973, USA}
\author{{Koichi Yamawaki}} \thanks{
      {\tt yamawaki@kmi.nagoya-u.ac.jp}}
      \affiliation{ Kobayashi-Maskawa Institute for the Origin of Particles and 
the Universe (KMI), Nagoya University, Nagoya 464-8602, Japan.}
\date{\today}

\begin{abstract} 
We find that the nonperturbative physics of the standard-model Higgs Lagrangian 
provides a dark matter candidate, ``dormant skyrmion in the standard model'', 
the same type of  the skyrmion, a soliton,  as in the hadron physics. 
It  is stabilized by another nonperturbative object in the standard model,  
the dynamical gauge boson of the hidden local symmetry, 
which is also an  analogue of  the rho meson. 
\end{abstract} 
\maketitle

\section{Introduction}

The standard model (SM) has been so successful that it is rather difficult to identify the
clue of the new physics beyond the SM, despite the most fundamental problem of the origin of mass, 
which might require physics beyond the SM~\footnote{For dynamical approach to this problem, 
see e.g., Ref.~\cite{Yamawaki:2016kdz} and references cited therein}. 
Moreover there is some concrete tension between the SM and the reality: 
apparent absence of the dark matter candidate, $\theta$ vacuum parameters due to instantons (strong CP problem, etc.),  absence of the first order phase transition for finite temperature
and  large enough CP violation required by the baryogenesis, etc.. 
Also, existence of the Landau pole invalidates the ``perturbative SM (pSM)'' 
(if not the SM itself)  at certain high energy. 
These possible failures of the SM may not exclude the SM but may only indicate our ignorance of 
the nonperturbative physics of the SM itself, 
although alternatively they could be solved in the context of the new physics beyond the SM.

Among those concrete problems,  the dark matter is a central mystery of the particle physics and the astrophysics today.

In this paper we demonstrate a possible resolution of the dark matter without explicit resource to the physics beyond SM,
namely, that  the {\it nonperturbative physics} of the SM Higgs Lagrangian 
provides a dormant dark matter candidate, the ``dormant skyrmion in the SM (DSSM)'', 
the same type of the soliton as the hadronic  skyrmion 
which already exists in the {\it  nonperturbative dynamics} of the nonlinear sigma model  
 (with Skyrme term) without  recourse to the underlying theory, QCD.

The hadronic skyrmion  is known to be 
stabilized by the  rho meson  as the gauge boson of the Hidden Local Symmetry (HLS) 
without ad hoc Skyrme term. 
{\it Such an HLS exists in any nonlinear sigma model}~\cite{Bando:1987br}. 
The SM Higgs Lagrangian is actually rewritten in the form of  nonlinear sigma model 
and hence has the HLS~\cite{Fukano:2015zua}. 
We show that the HLS gauge boson, ``SM-rho meson (SM$\rho$)'',  
acquires kinetic term by the {\it nonperturbative dynamics} of the SM  
and hence stabilizes the  DSSM. 
It provides a novel view of SM: the dark matter candidate 
{\it exists already inside the nonperturbative SM, though not the perturbative SM, 
without explicit recourse to beyond the SM.}

The SM Higgs Lagrangian is customarily written in the {\it linear} {\it sigma} {\it model}  
which is convenient for the perturbation theory. However, {\it pSM is not a whole story of the SM}, since there already exist sphaleron and instanton even for the weak coupling, 
which are well-known nonperturbative objects not to be described by the pSM. 
Also the 't Hooft-Polyakov monopole in the perturbatively renormalizable 
Georgi-Glashow model similar to the SM Higgs Lagrangian does exist 
even in the vanishing quartic coupling (BPS limit).  
Moreover, even in the perturbation the Higgs coupling grows indefinitely 
to hit the {\it Landau} {\it pole} in the ultraviolet region thus invalidating 
the perturbation itself. 
Thus in the SM as a full quantum theory the {\it nonperturbative} {\it effects} 
such as the bound states could emerge if not a narrow resonance before reaching 
the Landau pole without affecting the successful pSM at low energy. 
The nonperturbative quantum physics can often be better described 
by a different parameterization of the same Lagrangian at the classical level 
(e.g., see footnote \ref{CPN}), the {\it nonlinear} {\it sigma} {\it model} 
through the polar decomposition in the case at hand. 

In fact it was shown~\cite{Fukano:2015zua} that the SM Higgs Lagrangian on the broken vacuum 
can be straightforwardly cast into a scale-invariant version of the nonlinear sigma model 
based on the manifold $G/H=SU(2)_L\times SU(2)_R/SU(2)_V$, where 
{\it both the scale symmetry and chiral symmetry $G$ are realized nonlinearly}, with 
{\it the SM Higgs being nothing but a (pseudo) dilaton}, a Nambu-Goldstone (NG) boson of the spontaneously broken scale symmetry, 
together with the NG bosons of the spontaneous breaking of $G$ down to the subgroup $H$. 
(Though in that case the corresponding action is scale-invariant, 
just for convenience we shall hereafter call it the scale-invariant Lagrangian.)  
Once written in the form of nonlinear sigma model, 
one readily sees~\cite{Fukano:2015zua} that it has the HLS, 
since it is known~\cite{Bando:1984ej,Bando:1987br,Harada:2003jx} that 
any nonlinear sigma model is gauge equivalent to another model 
(HLS Lagrangian) 
having a symmetry $G_{\rm global}\times H_{\rm local}$, 
where $H_{\rm local}$ is a gauge symmetry (HLS), 
and the HLS gauge boson, SM$\rho$, is an auxiliary field as 
a static composite of the NG bosons.

Here we show that it actually develops  
kinetic term through {\it nonperturbative dynamics} 
of the NG bosons (longitudinal component of $W,Z$ bosons when the electroweak gauge switched on)
 {\it in the SM itself}, in a manner similar to the $CP^{N-1}$ model~\footnote{
The $CP^{N-1}$ is a nonlinear sigma model 
minimally written in terms of $2N$$-2$ NG bosons, but usually 
parameterized including redundancy: one constraint with Lagrange multiplier and 
the symmetry 
$SU(N)_{\rm global} \times$ $U(1)_{\rm local}$, 
where the redundant $U(1)_{\rm local}$ is nothing but the HLS whose gauge boson is
an auxiliary field to be solved away at the bare/perturbative theory. It is well established  \cite{Eichenherr:1978qa,Bando:1987br,Weinberg:1997rv,Harada:2003jx} that  in the large $N$ limit, the nonperturbative dynamics change the classical theory as if in  the broken phase into the unbroken 
phase 
in such a way that the HLS gauge boson necessarily 
acquires the kinetic term, becoming the true massless gauge boson, and the bare/perturbative NG bosons are no longer the NG bosons but have a mass, an extra free parameter given by the Lagrange multiplier. 
In $d$$=$$2$ dimensions  only the unbroken phase exists in conformity with the Mermin-Wagner-Coleman theorem,  while both phases exist in $2$$<$$d$$<$$4$ 
where the HLS gauge boson in the broken phase also acquires the kinetic term, 
and in addition a mass through the Higgs mechanism, precisely in the same way as the SM in the present paper. 
The $d$$=$$4$ model has a cutoff acting as a Landau pole where the induced kinetic term of HLS gauge boson vanishes, similarly to our present case
\cite{Bando:1987br,Harada:2003jx} (See also \cite{Weinberg:1997rv} for different formulation in $d$$=$$4$). 
Note that minimal parameterization of the {\it classical Lagrangian without redundancy  
is ill-defined at quantum level}~\cite{Bando:1987br}, 
thus the parameterization is crucial to  the nonperturbative quantum physics where the emergence of 
extra free parameters is mandatory. 
\label{CPN} 
}
as discussed long time ago~\cite{Bando:1987br,Harada:2003jx}. 
Once the HLS gauge boson acquired the 
kinetic term, it is well known~\cite{Igarashi:1985et} to 
stabilize the skyrmion, soliton in the nonlinear sigma model,  
without adding artificial 
Skyrme term by hand. 
It is also known~\cite{Park:2003sd,Kitano:2016ooc} that 
inclusion of the scalar meson, SM Higgs 
in our case, does not invalidate the stabilization.

Thus we find that the noperturbative SM  
provides a dark matter candidate, DSSM, 
with $I=J=0,1/2, \cdots$ (for isospin $I$ and spin $J$), 
living dormant {\it inside the SM but not beyond it}.  
Here we consider a complex-scalar bosonic-skyrmion, $X_s$,  with $I=J=0$.
The potential term of the SM Higgs Lagrangian, a small explicit breaking of the scale symmetry, 
which yields  the mass of the SM Higgs as a pseudo dilaton,
does not affect the essential features of the above skyrmion physics.

The salient feature of the DSSM is that its zero-momentum coupling to the SM Higgs 
as a pseudo-dilaton is {\it uniquely determined by the low energy theorem of the
spontaneously broken scale symmetry} for the dilaton-matter couplings as known 
for a long time in a different context~\cite{Carruthers:1971vz}. 
It yields  a unique constraint on its mass from the direct search experiments, 
i.e., $M_{X_s} \lesssim 13$ GeV 
from the LUX2016~\cite{Akerib:2015rjg,Akerib:2016vxi}. 
For such a low mass the SM Higgs can decay 
to ${\bar X}_s X_s$ and hence places 
the constraint 
$M_{X_s} \lesssim 18$ GeV from the invisible decay of the SM Higgs. 
The relic abundance is subject to 
the soliton size of DSSM, 
which we calculate in a benchmark parameter choice of HLS. 
The result for $M_{X_s} = {\cal O}(10)$ GeV is 
$\Omega_{X_s} h^2 \simeq 0.1$, 
consistently with the present data. 
We discuss it can be discovered/ruled out in the future experiments.

\section{SM Higgs Lagrangian as a scale-invariant HLS Lagrangian}

The SM Higgs Lagrangian 
takes the form of the $SU(2)_L\times SU(2)_R$ linear sigma model: 
\begin{eqnarray}
{\cal L}_{\rm SM}&=&
 |\partial_\mu h|^2 -\mu
 ^2 |h|^2 -\lambda|h|^4 
\nonumber
\\
&=&
\frac{1}{2} {\rm tr} \left( \partial_\mu M\partial^\mu M^\dagger \right)
 - \frac{\mu^2}{2} {\rm tr}\left(M M^\dagger\right)-\frac{\lambda}{4}  \left({\rm tr}\left(M M^\dagger\right)\right)^2 \,,
\label{Higgs1} 
  \end{eqnarray}
$ M\equiv
\frac{1}{\sqrt{2}}(i \tau_2 h^*, h)$ 
$=\frac{1}{\sqrt{2}}(\hat \sigma\cdot  {\bf 1}_{2\times 2} + 2 i \hat \pi) $, $
\left(\hat \pi \equiv  {\hat \pi}_a \frac{\tau_a}{2}\right)$.
Eq.(\ref{Higgs1})  is straightforwardly rewritten into the form~\cite{Fukano:2015zua}: 
    \begin{eqnarray}
{\cal L}_{\rm SM} 
&= &
\frac{1}{2} \left(\partial_\mu \sigma \right)^2  +\frac{\sigma^2}{4}{\rm tr} \left(\partial_\mu U \partial^\mu U^\dagger\right) -V(\phi) 
=\chi^2 \cdot \left[ \frac{1}{2} \left(\partial_\mu \phi\right)^2  +\frac{v^2}{4}{\rm tr} \left(\partial_\mu U \partial^\mu U^\dagger\right)\right] -V(\phi)\,, 
  \label{Higgs2}  \\
  V(\phi) 
  & \equiv& 
  \frac{\mu^2}{2} {\rm tr}\left(M M^\dagger\right) +\frac{\lambda}{4}  \left({\rm tr}\left(M M^\dagger\right)\right)^2 
  =\frac{\mu^2}{2} \sigma^2 + \frac{\lambda}{4} \sigma^4
=\frac{\lambda}{4} v^4 \left[\left(\chi^2 -1\right)^2-1\right]
 \,,  \\
  \nonumber
      \end{eqnarray}
where we have used the standard  polar decomposition~\footnote{
See e.g., section 5.3 in Ref.~\cite{Bando:1987br}.
Note that the Cartesian coordinate $(\hat \sigma, \hat\pi)$ are {\it chiral non-singlet} transforming into each other under the chiral transformation $g_{L/R} $$\in $$SU(2)_{L/R}$. 
In the polar decomposition, on the other hand,  the chiral transformation  of $M$, $M\rightarrow g_L \, M\, g_R^\dagger$, is inherited by $U$ consisting of the 
genuine angular coordinates  $\pi$ (NG bosons, becoming totally the {\it electroweak gauge parameters} to be absorbed into $W,Z$ bosons), $U\rightarrow g_L \, U\, g_R^\dagger$, 
 while the {\it genuine radial component} $H$
  is a {\it chiral singlet (electroweak gauge singlet)}, $ H $ $\rightarrow$ $ H$, and
 so is the  physical  mode, the SM Higgs field $\phi$, which is {\it electroweak-gauge invariant} as it should be. 
 The physical SM Higgs field  $\phi$ is actually the {\it dilaton} transforming as $\delta_D \phi(x)=
v + x^\mu \partial_\mu \phi(x)$ under the scale transformation. Note also that the chiral singlet potential term $V(\phi)$, having only $\phi$ and no NG bosons $\pi$, breaks the scale symmetry explicitly by the amount $\lambda$.
} 
of the $2 \times 2$ complex matrix $M$ into  a positive
Hermitian matrix $H$ times  a unitary matrix $U$: 
\begin{eqnarray}
M(x) & = & H(x)\cdot U(x)\,,   
\nonumber\\
H(x)&=&\frac{ {\bf 1}_{2\times 2}}{\sqrt{2}} \cdot \sigma(x)\,\,, \quad
\left(\sigma(x) \equiv \sqrt{
{\hat \sigma}^2+{\hat \pi}^2} 
=v \cdot \exp \left(\frac{\phi(x)}{v}\right)=v\cdot \chi(x) 
\right)\,,
\nonumber\\
U(x)&=& \exp \left( \frac{2 i \pi(x)}{v}\right) \,, \quad  \left(\pi \equiv  \pi_a \frac{\tau_a}{2}\right) \,,
\label{polar}
 \end{eqnarray}
with 
$\langle \sigma(x)\rangle 
= v = \sqrt{\frac{-\mu
^2}{\lambda}} \ne 0$ 
($\langle \chi(x)\rangle=1\,, \langle \phi(x) \rangle = 0$).

Thus the SM Higgs Lagrangian Eq.(\ref{Higgs1})  
is trivially  identical to Eq.(\ref{Higgs2}). Note that the kinetic term of the latter, 
$\chi^2 \cdot \left[ \frac{1}{2} \left(\partial_\mu \phi\right)^2  +\frac{v^2}{4}{\rm tr} \left(\partial_\mu U \partial^\mu U^\dagger\right)\right] $
 contains the usual nonlinear sigma model 
 $\frac{v^2}{4}{\rm tr} \left(\partial_\mu U \partial^\mu U^\dagger\right) $ 
 which transforms as dimension 2 to make the action not scale-invariant.
However, the extra dilaton factor 
$\chi^2(x)=e^{2\phi(x)/v}$,  
transforming as dimension 2, makes the whole kinetic term to be dimension 4. Hence 
the action becomes scale-invariant as it should, since it is  just a rewriting of the original kinetic term in Eq.(\ref{Higgs1}) which is scale-invariant (dimension 4).

Actually the kinetic term coincides with the 
scale-invariant nonlinear chiral Lagrangian based on the coset $G/H=SU(2)_L\times SU(2)_R/SU(2)_{L+R}$, 
with  the scale symmetry as well as the chiral symmetry being realized nonlinearly. 
Thus  : 
{\it the SM Higgs 
$\phi$ is nothing but a pseudo dilaton, 
with 
the explicit breaking of the scale symmetry from the potential term $V(\phi)$ 
characterized by 
\begin{equation}
\lambda=\frac{M_\phi^2}{
2 v^2}
\simeq 
\frac{(125 \,{\rm GeV})^2}
{
2\times (246\, {\rm GeV})^2
} \simeq \frac{1}{
8} \ll 1\,,
\end{equation}
which is very close to the ``conformal limit'', 
$\lambda\rightarrow 0$ with $v=$ fixed}  ($V(\phi)\rightarrow 0$)~\footnote{
The opposite  limit, $\lambda$ $\rightarrow$$\infty$ with $v$$=$ fixed, leads to the 
ordinary nonlinear sigma model, where we also have $V(\phi) \rightarrow 0$ but 
with $\chi(x) \equiv 1$, so that the scale symmetry compensated by $\chi^2$ factor 
in Eq.(\ref{Higgs2}) is completely lost. See Ref.\cite{Fukano:2015zua}. 
Either limit has no
 $\lambda$ coupling, but has derivative couplings instead, which are ``weak'' in the low energy
 $p^2/(4\pi v)^2 \ll 1$, so that the perturbation according to the derivative expansion (``chiral perturbation theory'') makes sense.
},
the limit  
corresponding to the ``Bogomol'nyi-Prasad-Sommerfield (BPS) limit'' of 
't Hooft-Polyakov monopole in the Georgi-Glashow model~\cite{Harvey:1996ur}, 
similarly to the SUSY flat direction~\footnote{
Even if we take such a conformal/BPS limit, 
the theory is still an interacting theory with {\it derivative} {\it coupling} as in the usual chiral Lagrangian,
 and thus the quantum corrections will produce the trace anomaly of dimension 4, 
$\sim$ $v^4$ $ \chi^4 \ln \chi$, as a new source of the
SM Higgs mass as a pseudo-dilaton, which, however, 
would do not affect the basic nature of the DSSM property discussed here, similarly to the tiny explicit 
scale-symmetry breaking 
 in the tree-level potential $V(\phi)$ with 
$\lambda$ $\simeq $ $1/8$ $\ll$ $1$. 
}.

Now that the SM Higgs Lagrangian Eq.(\ref{Higgs1}) is rewritten in the form of the nonlinear sigma model, Eq.(\ref{Higgs2}),
 we can further rewrite Eq.(\ref{Higgs2}) into another
gauge equivalent form having redundant gauge symmetry in terms of the gauge fields as auxiliary fields~\cite{Fukano:2015zua}: it is well-known~\cite{Bando:1984ej,Bando:1987br,Harada:2003jx} that nonlinear sigma model on the manifold 
$G/H$ is gauge equivalent to the Hidden Local Symmetry (HLS) model having the internal symmetry $G_{\rm global} \times H_{\rm local}$ 
with $H_{\rm local}$ being the HLS.

In the case at hand, based on the manifold $G/H=SU(2)_L \times SU(2)_R/SU(2)_{L+R}$, 
 the HLS $H_{\rm local}= [SU(2)_{L+R}]_{\rm local}$was introduced as  a redundancy of dividing $U$ into two parts:  
$U = e^{2 i\pi/v}
= \xi_L^\dagger\cdot \xi_R$ such that $\xi_{L/R}$ transform as 
$\xi_{L/R} (x)\rightarrow h(x)\cdot \xi_{L/R}(x) \cdot {\hat g}_{L/R}^\dagger$ and 
so do the covariant derivatives $ D_\mu \xi_{L,R}(x)\equiv (\partial_\mu -i \rho_\mu(x)) \xi_{L.R}(x)$, where 
$h(x) \in H_{\rm local}$ and ${\hat g}_{L/R}\in G_{\rm global}$, 
$\rho_\mu(x)$ being the gauge boson of $H_{\rm local}$. 
We may parametrize $\xi_{L,R}=e^{i {\bar \rho}/F_\rho}\cdot e^{\mp i \pi/v}$, with ${\bar \rho} (x)$ being the fictitious NG boson to be absorbed into $\rho_\mu(x)$, and $F_\rho$ its decay constant  (identical to the conventional coupling of $\rho_\mu$ to the vector current). 
When we fix the gauge of HLS as $\xi_L^\dagger=\xi_R=\xi=e^{i \pi/v}$ 
(unitary gauge ${\bar \rho}(x)=0$) 
such that $U$ $=$ $\xi^2$, 
$H_{\rm local}$ and $H_{\rm global}$ $(\subset G_{\rm global})$  
get simultaneously broken spontaneously (Higgs mechanism), leaving the 
diagonal subgroup $H=H_{\rm local} + H_{\rm global}$, 
which is nothing but the subgroup $H$ of the original  $G$ for $G/H$: 
$H$ $\subset$ $G$. Then the theory with 
the symmetry $G_{\rm global}$ $ \times$ $ H_{\rm local}$ is gauge equivalent to 
that on the manifold $G/H$. 
See e,g., Refs.~\cite{Bando:1987br,Harada:2003jx}.

We here define Maurer-Cartan 1-forms $\alpha_{\mu,R,L}\equiv \frac{1}{i}\partial_\mu \xi_{R,L} \cdot \xi_{R,L}^\dagger$ 
and the covariantized ones ${\hat \alpha}_{\mu,R,L}\equiv \frac{1}{i}D_\mu \xi_{R,L}\cdot \xi_{R,L}^\dagger =
 \alpha_{\mu,R,L}- \rho_\mu$,
which transform 
 as ${\hat \alpha}_{\mu, R,L} \rightarrow h(x){\hat \alpha}_{\mu, R,L} h(x)^\dagger$.
We further take the
 linear combinations $ {\hat \alpha}_{\mu,||}=\frac{1}{2} ({\hat \alpha}_{\mu,R} + {\hat \alpha}_{\mu,L})=\frac{1}{2} ({\alpha}_{\mu,R} + {\alpha}_{\mu,L})- \rho_\mu=\left[\frac{1}{F_\rho} \partial_\mu \bar \rho
  - \frac{i}{2 v^2}[\partial_\mu \pi,\pi] +\cdots\right] - \rho_\mu$ and 
 ${\hat \alpha}_{\mu,\perp}=\frac{1}{2} ({\hat \alpha}_{\mu,R} - {\hat \alpha}_{\mu,L})=\frac{1}{2} ({\alpha}_{\mu,R} -{\alpha}_{\mu,L})=\frac{1}{2i}\xi_L\cdot \partial_\mu U\cdot \xi_R^\dagger=\frac{1}{2i}\xi_R \partial_\mu U^\dagger\cdot \xi_L^\dagger\,,
 $. 
 Thus the  HLS Lagrangian consists of the two $G_{\rm global} \times H_{\rm local}$-invariants: 
 \begin{eqnarray}
{\cal L}_{\rm HLS} &=& 
v^2 \,{\rm tr}[\hat{\alpha}_{\mu,\perp}^2] + a v^2 \,{\rm tr}[\hat{\alpha}_{\mu,||}^2] \,, 
\label{HLS}
\\
 && 
 v^2\, {\rm tr}[\hat{\alpha}_{\mu,\perp}^2] =v^2\, {\rm tr}[\alpha_{\mu,\perp}^2] =\frac{v^2}{4}\, {\rm tr} \left(\partial_\mu U \partial^\mu U^\dagger\right)\,, 
 \nonumber \\ 
&&
a v^2 {\rm tr}[\hat{\alpha}_{\mu,||}^2]=a v^2 {\rm tr}[\rho_\mu - \alpha_{\mu,||}]^2 
=    F_\rho^2 \,{\rm tr}\left[
 \left(\rho_\mu - \frac{1}{F_\rho} \partial_\mu \bar \rho \right) 
 + \frac{i}{2 v^2}[\partial_\mu \pi,\pi] +\cdots
 \right]^2 \,, 
 \nonumber 
  \end{eqnarray}
where $F_\rho^2=a v^2$ to normalize the kinetic term of the fictitious NG boson ${\bar \rho}(x)$. The first term is identical to the original nonlinear sigma model, while
the second term is the HLS-invariant mass term of the $\rho_\mu$ as obvious in the unitary gauge $\bar \rho  =0$, which also contains the $\rho\pi\pi$ coupling.

Without kinetic term of the auxiliary field $\rho_\mu$, 
the second term in Eq.(\ref{HLS})  
vanishes: $a v^2 {\rm tr}[\rho_\mu - \alpha_{\mu,||}]^2=0$, 
when equation of motion $\rho_\mu=\alpha_{\mu,||}$ is used. 
Then Eq.(\ref{HLS}) is simply reduced back to the original nonlinear sigma model 
as it should be for the auxiliary field $\rho_\mu$.

It was further shown~\cite{Fukano:2015zua} that 
the SM Higgs Lagrangian Eq.(\ref{Higgs2}) (hence Eq.(\ref{Higgs1}) as well)  is gauge equivalent to the scale-invariant 
version~\cite{Kurachi:2014qma} 
of the HLS Lagrangian 
Eq.(\ref{HLS})~\cite{Bando:1984ej,Bando:1987br,Harada:2003jx}, 
up to  the scale-violating potential $V(\phi)$, 
for the internal symmetry 
$G_{\rm global} \times H_{\rm local}
=[SU(2)_L \times SU(2)_R]_{\rm global}\times [SU(2)_{L+R}]_{\rm hidden}$.
Having extra dilaton kinetic term and the dilaton factor $\chi^2=e^{2\phi/v}$ responsible for  the theory to be scale-invariant (up to $V(\phi)$), it reads in the unitary gauge:
\begin{eqnarray}
 {\cal L}_{\rm SM/HLS}
 &=&
  \,\chi^2 \cdot 
\left( 
\frac{1}{2} (\partial_\mu \phi)^2 +  v^2\, {\rm tr}[\hat{\alpha}_{\mu,\perp}^2] + a v^2 \, {\rm tr}[\hat{\alpha}_{\mu,||}^2] 
 \right) -V(\phi)
 \nonumber\\ 
 &=&
 e^{\frac{2\phi}{v}} \left[\frac{1}{2} (\partial_\mu \phi)^2 
 + \frac{v^2}{4} {\rm tr}(\partial_\mu U \partial^\mu U^\dagger) 
 + a\, v^2 \,{\rm tr} \left(\rho_\mu -\frac{i}{2 v^2}[\partial_\mu \pi, \pi]
 +\cdots\right)^2\right]  - V(\phi)\,,
  \label{SHLS}
\end{eqnarray} 
where the term 
$\chi^2 a  v^2 {\rm tr}[\hat{\alpha}_{\mu,||}^2] 
=\chi^2 a  v^2 {\rm tr}[\rho_\mu - \alpha_{\mu,||}]^2$ 
is the $G$-invariant/scale-invariant mass term of  
the  SM-rho meson (SM$\rho$).

 Now we discuss that the kinetic term of SM$\rho$ 
 is generated dynamically by the quantum loop~\cite{Bando:1987br,Harada:2003jx}, 
similarly to the $CP^{N-1}$ model (See footnote \ref{CPN}), 
in the same sense as the dynamical generation of the 
 kinetic term (and the quartic coupling as well) of the composite Higgs 
in the Nambu-Jona-Lasinio model, which is an auxiliary field at the tree level or 
at composite scale (Landau pole of the pSM)~\cite{Eguchi:1976iz}. 
 In order to discuss the off-shell $\rho_\mu$ (in space-like momentum) relevant to 
the skyrmion stabilization~\footnote{
The $M_\rho^2$ would develop (potentially large) imaginary parts  
in the time-like region for decaying to the $WW, WZ$ if $M_\rho$ $>$ $ 2 M_{W/Z}$. 
However, this would not affect the skyrmion physics which is relevant to the
 space-like $\rho$. 
}, 
we adopt the background field gauge as in Ref.~\cite{Harada:2003jx}.
 Integrating out  
 high frequency  modes  from the composite scale  
$\Lambda$   
to the scale $\mu$ in the Wilsonian sense at one-loop, 
the kinetic term is 
generated with the gauge coupling $g$~\cite{Harada:2003jx}: 
\begin{equation} 
- \frac{1}{2g^2}  {\rm tr} [\rho_{\mu\nu}^2] 
\,,\qquad 
\frac{1}{g^2}=
 \frac{C_2(G)}{(4\pi)^2}\, \left(\frac{a^2}{24}+ \left(-\frac{11}{3}+\frac{1}{24}\right)\right)\, \ln \frac{\Lambda^2}{\mu^2}
\,, 
\end{equation} 
with $C_2(G)=N_f$ and the number of chiral flavors $N_f=2$, 
where in the second equation the first term is the loop contribution of $\pi$ (longitudinal $W/Z$ when the electroweak gauging switched on) with $g_{\rho\pi\pi}=a/2$, 
the second the  loop of the dynamically generated SM rho, with the usual factor $(-11/3)$ 
and the last one is from the loop of the would-be NG boson  (the longitudinal 
  SM rho) having the $\rho$ coupling $1/2$, which ends up with 
 $ \frac{1}{g^2}= \frac{N_f}{(4\pi)^2}\frac{a^2-87}{24}\ln \frac{\Lambda^2}{\mu^2}$. For $a>\sqrt{87}$ and/or $\mu^2 >M_\rho^2$ 
  (second and third terms are decoupled),
  we find $1/g^2 \rightarrow 0$ for $\mu\rightarrow \Lambda$, 
which allows us to identify the scale $\Lambda$ 
with the Landau pole. 

In the present paper, therefore, 
we confine ourselves only to the case $a\gg \sqrt{87}$~\footnote{
If the SM is regarded as an effective theory of some underlying theory to provide the bare $\rho$ kinetic term already at $\Lambda$, the parameter $a$ could be
much smaller  as the QCD rho meson with $a\simeq 2$.}.     
Rescaling the kinetic term to the canonical one, we have the (off-shell) mass 
$M_\rho^2(\mu) =a(\mu) g^2(\mu) v^2 \searrow$ as $\mu/\Lambda \searrow$, 
where $a(\Lambda)=a$. 
(As shown in Ref.~\cite{Harada:2003jx}, 
the parameter $a$ also gets corrected by the HLS gauge loop contributions,
 to have the renormalization-group running 
when it scales down from the Landau pole $\Lambda$ to the infrared scale $\mu$. 
More explicit discussion based on the renormalization group analysis 
will be presented in another publication.)

Note that the one-loop contribution to the kinetic term of the auxiliary field is not literally one-loop perturbation but 
  actually corresponds to an
   infinite geometric summation of the one-loop diagram (``bubble sum'') in the explicit nonperturbative treatment without auxiliary field in $1/N$ 
expansion~\cite{Eguchi:1976iz}. 
(Also see the large $N$ description for the $CP^{N-1}$ model in the 
footnote~\ref{CPN}. 
In the present SM case, the number of the chiral flavors $N_f(=2)$ 
would play the role of the $N$ for the $CP^{N-1}$ model. 
More explicit derivation is to be supplied in another publication.)

\section{Emergence of the DSSM  from the SM Higgs Lagrangian}

Now that the SM-rho meson has been dynamically generated,
the Lagrangian of the SM with the HLS, Eq.(\ref{SHLS}), thus takes the form~\cite{Fukano:2015zua}:  
\begin{eqnarray} 
{\cal L} 
&=& 
\chi^2 \cdot 
\left( 
\frac{1}{2} (\partial_\mu \phi)^2 +   
 \frac{v^2}{4}{\rm tr}(\partial_\mu U \partial^\mu U^\dagger)
 \right)
  + 
  \chi^2 \cdot a\, v^2 \,{\rm tr} 
 \left(
g\cdot \rho_\mu -\frac{i}{2 v^2} 
 [\partial_\mu \pi, \pi]+\cdots
 \right)^2
- \frac{1}{2} {\rm tr}[
\rho_{\mu\nu}^2] 
+ \cdots 
\,, \label{Lag}
\end{eqnarray} 
where the SM$\rho$ kinetic term is re-scaled to the canonical one, i.e.,  $\rho_\mu(x) \rightarrow g \cdot \rho_\mu(x)$, and the last  ``$+ \cdots$'' includes the scale-symmetry violating potential term $V(\phi)$ and quantum mechanically induced higher terms. 
This is our basic Lagrangian for the SM Higgs including the nonperturbative quantum effects. 
Near the Landau pole, such that $g \gg 1$ with $a=$ fixed, this is reduced  to the original  SM Higgs Lagrangian at classical level, Eq.(\ref{Higgs2}) (equivalently Eq.(\ref{Higgs1})).

The soliton energy $E$ thus takes the form similar to that 
analyzed in Ref.\cite{Park:2003sd} based on the scale-invariant version of the 
HLS Lagrangian for QCD hadrons such as 
$\pi,\rho,\omega,\sigma$ (except for the potential term as well as
the pion mass  and $\omega$ meson terms which are missing in our case):
\begin{eqnarray}
E &=& E_\pi + E_{\pi\rho} + E_\rho + E_\chi,  \notag \\
E_\pi &=& 4\pi \int_0^\infty r^2dr  v^2 \chi(r)^2 \left( \frac{F'(r)^2}{2} +\frac{\sin^2{F(r)}}{r^2} \right), \nonumber \\  
E_{\pi\rho} &=& 4\pi \int_0^\infty dr a v^2 \chi(r)^2 \left( G(r)+1-\cos{F(r)} \right), 
\nonumber \\ 
E_\rho &=& 4\pi \int_0^\infty dr \frac{1}{g^2} \left( G'(r)^2 +\frac{G(r)^2(G(r)+2)^2}{2r^2} \right), 
\nonumber \\  
E_\chi &=& 4\pi \int_0^\infty r^2dr  \frac{v^2}{2} \left( \chi(r)^2 \phi'(r)^2 + \frac{M_\phi^2}{4}(\chi(r)^2-1)^2 \right), 
\label{solitoneq1}
\end{eqnarray}
where   
we took the usual hedgehog ansatz for the  fields 
as done in Ref.\cite{Park:2003sd}: 
$\chi=e^{\phi(r)}, \rho_{\mu=i}^a=\epsilon^{ika}\hat{r}^k\frac{G(r)}{gr}, 
\rho_{\mu=i}=0, U=e^{i\vec{\tau}\cdot \hat{r} F(r)}$
with $\hat{r}$ being unit normalized vector. 
In the above expressions the prime attached on the fields denotes the derivative with respective to $r$. 
The equations of motion for $F(r), \phi(r)$, and $G(r)$ are obtained 
by minimizing the soliton energy: 
\begin{eqnarray}
F''(r) &=& -\left( \frac{2}{r}+\phi'(r) \right) F'(r) + \frac{1}{r^2} 
\left( 2a (G(r)+1) \sin{F(r)}+(1-a) \sin{2F(r)} \right), \notag \\
G''(r) &=& a (gv)^2 (G(r)+1-\cos{F(r)} ) + \frac{G(r)(G(r)+1)(G(r)+2)}{r^2},\nonumber  \\
\phi''(r) &=& -\phi'(r)^2-\frac{2}{r} \phi'(r)+\frac{M_\phi^2}{2}(\chi(r)^2-1) + F'(r)^2 + \frac{2}{r^2}\sin^2{F(r)} . 
\label{soliton}
\end{eqnarray}
To obtain a solution for the topological number $N_{X_s}=1$, we take $F(0)=\pi$. 
From the equation of motion one can see the boundary condition for $G(0)$ as 
$G(0)=-2$. 
The profile functions should also satisfy the boundary conditions at infinity as 
\begin{eqnarray}
F(\infty) =0, \ \ G(\infty)=0, \ \ \chi(\infty)=1. 
\label{solitoneq12}
\end{eqnarray}

It is also well known~\cite{Igarashi:1985et,Park:2003sd} that this system without the Skyrme term has a stabilized skyrmion, 
irrespectively of the  factor $\chi^2$ which makes  the rho and the skyrmion as well as the Higgs to be scale-invariant nonlinearly  
 (up to small effects of the term $+ \cdots$ in Eq.(\ref{Lag})).

In fact the SM$\rho$ kinetic term is reduced to the Skyrme term in  
the limit 
$a \to \infty$ with $g$ fixed~\cite{Igarashi:1985et} 
(infinite mass limit $M_\rho^2=a g^2 v^2 \rightarrow \infty$), since 
the SM$\rho$ field $\rho_\mu$ can be integrated out via the equation of motion, 
$g\cdot \rho_\mu \approx \alpha_{\mu||}$, in such a way that  
$g^2 \cdot \rho_{\mu\nu} \approx i  [\hat\alpha_{\mu, \perp}, \hat{\alpha}_{\nu, \perp}]
= i  [\alpha_{\mu, \perp}, \alpha_{\nu, \perp}]$,
which is plugged back into the Lagrangian Eq.(\ref{Lag}). 
In this limit the
SM$\rho$  effects other than the DSSM are invisible in the collider physics, so that {\it all the successful results of the pSM are intact}. 
In fact the perturbation theory of the SM Higgs in Eq.(1) is 
independent of the parameterization~\cite{Alonso:2016oah}, 
and is exactly the same as Eq.(2) since the auxiliary field
has no effect in the pSM. 
The SM$\rho$ kinetic term now reads
the Skyrme term~\cite{Skyrme:1961vq}: 
\begin{equation}   
- \frac{1}{2} {\rm tr}[
\rho_{\mu\nu}^2]
\Bigg|_{g={\rm fixed}} 
\stackrel{a\rightarrow \infty}{\longrightarrow}  -\frac{1}{2\, g^2} {\rm tr} \left[[\alpha_{\mu,\perp},\alpha_{\nu,\perp}]\right]^2
=
 \frac{1}{32 g^2} {\rm tr}[
 [ \partial_\mu U U^\dag, \partial_\nu UU^\dag ]^2
]
\,, \label{Skyrmi-term}
\end{equation}
where we used $\hat{\alpha}_{\perp \mu}=\xi_L (\partial_\mu U) \xi_R^\dag/(2i)$. 
Note that the ``Skyrme term limit'', $a\rightarrow \infty$ with $g=$ fixed, is different from the ``classical limit'', $g\rightarrow \infty$ with $a=$ fixed, which we already discussed is the limit going back to the original SM Higgs Lagrangian at classical level, Eq(\ref{Higgs1}) and/or Eq.(\ref{Higgs2}).

More explicitly, from Eq.(\ref{soliton}) in the limit $a\rightarrow \infty$ with $g=$ fixed, 
$G(r)$ can be reduced to $G(r) = \cos{F(r)}-1$ in the background dilaton profile, 
then $E_\rho$ becomes the Skyrme term. 
 (One can easily see that the energy $E_\rho$ with $G(r) = \cos{F(r)}-1$ is equivalent to 
the one given the Skyrme term Eq.~(\ref{Skyrmi-term}) when 
the hedgehog ansatz $U=e^{i\vec{\tau}\cdot \hat{r} F(r)}$ is assumed.) 
Then under minimization of  the new soliton energy 
the equations of motion reads 
\begin{eqnarray}
F''(r) &=& \frac{-1}{r^2\chi(r)^2+2(gv)^2\sin^2{F(r)}} \notag \\ 
&&
\times \left( 
\chi(r)^2(-\sin{2F(r)}+2rF'(r))+2r^2 \chi(r)^2\phi'(r)F'(r) +\frac{\sin{2F(r)}}{(gv)^2} 
\left( F'(r)^2 -\frac{\sin^2{F(r)}}{r^2} \right)
\right), \notag \\
\phi''(r) &=& -\phi'(r)^2 -\frac{2}{r}\phi'(r)+\frac{M_\phi^2}{2}(\chi(r)^2-1)+ F'(r)^2 + \frac{2}{r^2}\sin^2{F(r)}\,. 
\label{solitoneq2}
\end{eqnarray}
This happen to be similar to the soliton equations analyzed in Ref.\cite{Kitano:2016ooc} up to the difference in the potential term (higher terms like $\phi^3, \phi^4\cdots $).
While 
the Skyrme term in Ref.\cite{Kitano:2016ooc}  was introduced by hand {\it from outside of the SM}, in our case the same Skyrme term is {\it generated by the
nonperturbative dynamics of the SM itself}. 
Note that the hedgehog profile of the Higgs field takes the form somewhat different between the dilaton in our case and  the scale non-invariant shifted field in Ref.\cite{Kitano:2016ooc}, but both cases numerically lead to similar results. 
Moreover, the crucial difference is that our Higgs and DSSM are both dictated by the scale symmetry, while those of Ref.\cite{Kitano:2016ooc} are completely free parameters. (See the discussions below through sections \ref{DTE} and \ref{IDTE}, to be given before the explicit discussions on the soliton properties of the DSSM in section \ref{thermal}.)

Thus the DSSM  emerges as a soliton solution from the Lagrangian Eq.(\ref{Lag}) (here we consider
$I=J=0$)
to be a topological bosonic matter carrying the topological number, which we call $U(1)_{X_s}$. 
The $U(1)_{X_s}$ symmetry protects the decay of the DSSM 
($X_s$) completely, so the $X_s$ can be a dark matter candidate.

The DSSM is essentially generated by the scale-invariant part of Eq.(\ref{Lag}) and hence 
its coupling is dictated by the nonlinear realization of the scale symmetry. 
This yields a salient feature of the DSSM which characterizes phenomenological consequences as the constraints from 
both direct and indirect detection experiments to be discussed in sections \ref{DTE} and \ref{IDTE}.

 In the low energy limit
$q^2 \ll v^2 \simeq (246 \, {\rm GeV})^2$ 
the DSSM ($X_s$)  coupling to the SM Higgs $\phi$ as a pseudo dilaton is  
unambiguously determined by the low-energy theorem of the scale symmetry~\cite{Carruthers:1971vz}~\footnote{The low-energy theorem for the dilaton $\phi(q_\mu)$ coupling to matter 
$B$ at $q_\mu$ $\rightarrow$ $ 0$ reads $g_{\phi B B}$ $=$ $2M_B^2/v$, 
$M_B/v$ for complex scalar and spin $1/2$ fermion, 
respectively \cite{Carruthers:1971vz},  
which is consistent with the scale invariance of the mass term;
 $M_B^2 B^\dagger \chi^2 B$ $=$ $M_B^2 B^\dagger B $ $+$ $(2 M_B^2/v) $ $\phi B^\dagger B$ $+\cdots$ 
and  
 $M_B \bar{B} \chi B$ $=$ $M_B \bar{B} B + (M_B/v) \phi$ $\bar{B} B$ $+\cdots$ 
 for complex scalar and spin $1/2$ fermion, 
respectively. 
Incidentally, this also applies to the Yukawa coupling  of SM Higgs $\phi$ to other matter, 
the quarks/leptons, $g_Y^{q/l}$ $=$ $M_{q/l}/v$, 
which is consistent with the SM Higgs being a pseudo-dilaton. }, 
as described 
by the scale-invariant form of the lowest derivative effective Lagrangian: 
\begin{equation}
 {\cal L}_{X_s}(q \ll v) = 
\partial_\mu {X_s}^\dag \partial^\mu X_s - M_{X_s}^2 \chi^2 X_s^\dag X_s  
\,. 
\end{equation}  
From this we can readily read the 
$\phi-X_s$ interaction: 
\begin{equation} 
{\cal L}_{X_s}^{\rm int}(q \ll v) = 
- \frac{2M_{X_s}^2}{v^2} \left( 
v \cdot \phi X_s^\dag X_s +  \phi^2 X_s^\dag X_s + \cdots 
\right) 
\,, 
\end{equation} 
the first term of which is relevant to the dark matter detection experiments for 
weakly-interacting massive-particle (WIMP): 
\begin{equation} 
{\cal L}_{\rm int}^{\phi X_sX_s} (q \ll v) 
= - 2 \lambda_{X_s} v \cdot (\phi X_s^\dag X_s)  
 \,,\,  
 \lambda_{X_s} \equiv \frac{M_{X_s}^2}{v^2}
 \,. \label{3-point}
\end{equation}

\section{DSSM in direct detection experiments} 
\label{DTE}

The DSSM $X_s$ can be measured in the direct detection experiments of dark matter 
such as the LUX and PandaX-II experiments~\cite{Akerib:2015rjg,Akerib:2016vxi}. 
The spin-independent (SI) elastic scattering cross section of the DSSM  
$X_s$ per nucleon $(N=p,n)$,
through the Higgs $(\phi)$ exchange at zero-momentum transfer, 
can be calculated by the standard formula  as a function of the DSSM mass $M_{X_s}$:  
\begin{eqnarray} 
&& 
\sigma_{\rm SI}^{\rm elastic/nucleon} (X_s N \to X_s N) 
= 
\frac{\lambda_{X_s}^2}{\pi M_\phi^4}  
\left[ 
\frac{Z}{A} \cdot m_*(p, X_s) g_{\phi pp} \frac{m_p}{M_{X_s}} 
+ \frac{A-Z}{A} \cdot m_*(n,X_s) 
g_{\phi nn}  \frac{m_n}{M_{X_s}}
 \right]^2  
 \,,
\end{eqnarray}    
with the target nucleus Xe: ($Z=54$, $A=131.293$, $u=0.931$ GeV),
$m_{p(n)} \simeq $ 938(940) MeV, 
$v \simeq 246$ GeV and  
$M_\phi \simeq 125$ GeV, 
where $ m_*(N,X_s) = \frac{M_{X_s} m_N}{M_{X_s} + m_N} $ and  
$ g_{\phi pp(nn)} 
= 
\sum_q \sigma_q^{(p(n))}/v  
\simeq 
0.248(0.254) $~\cite{Ohki:2008ff,Hisano:2015rsa}.

  \begin{figure}[t]
\begin{center}
   \includegraphics[scale=0.65]{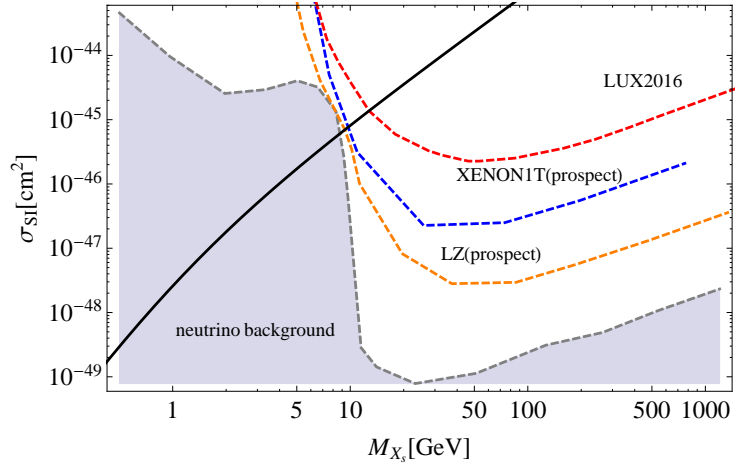} 
\caption{ 
The spin-independent elastic scattering cross section of the DSSM  
$X_s$ per nucleon as 
a function of the mass $M_{X_s}$ in unit of ${\rm cm}^2$ (solid curve). 
Also have been shown the most stringent constraint at present from the latest 
LUX2016 experiment~\cite{Akerib:2016vxi} 
and projected experiments with the xenon target by the end of this 
decade~\cite{Feng:2014uja}.  
The gray domain, surrounded by the dashed curve on the bottom, 
stands for 
the atmospheric and astrophysical neutrino background~\cite{Billard:2013qya}.   
}
\label{mB-sigma}
\end{center} 
 \end{figure}

The plot is shown in Fig.~\ref{mB-sigma} along with the currently strongest exclusion limit from the 
latest LUX2016 experiment~\cite{Akerib:2016vxi}.   
The present upper bound on the 
DSSM  mass $M_{X_s}$ 
is thus read off as  
\begin{equation} 
 M_{X_s} \lesssim 13 \,{\rm GeV} 
 \,.  \label{mB:limit:LUX}
\end{equation} 
Note that we have the {\it upper bound instead of the lower bound} in contrast to 
conventional WIMP models due to the characteristic dilatonic coupling proportional 
to $m_{X_s}^2$ as in Eq.(\ref{3-point}).
In Fig.~\ref{mB-sigma} we have also shown 
the expected limits from the projected  
direct detection experiments with the target nucleus of 
xenon~\cite{Feng:2014uja} which will be activated by the end of this decade. 
From this, one can see that the XENON1T experiment has the sensitivity 
to exclude, or discover the 
DSSM with the mass up to $\simeq$ 10 GeV, 
and it will get lower up to $\simeq 9$ GeV 
for the LUX-ZEPLIN (LZ) experiment.

\section{Indirect search limit from Higgs invisible decay}  
\label{IDTE}

\begin{figure}[t]
\begin{center}
   \includegraphics[scale=0.55]{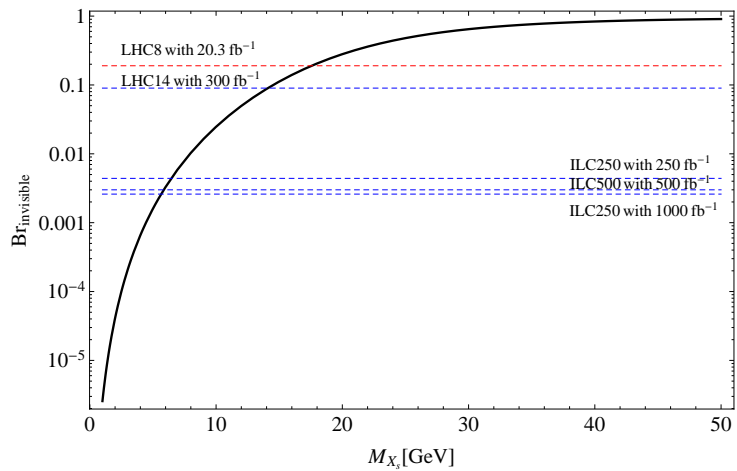} 
\caption{ 
The branching fraction of the Higgs decaying to the DSSM pair as 
a function of the mass $M_{X_s}$ (solid curve). 
Also have been plotted the most stringent (95\% C.L.) 
constraint at present from the  
LHC run I combined with the early stage of the 13 TeV run  
reported by the CMS group~\cite{CMS:2016rfr}, 
and the expected 95\% C.L. limits in the projected collider experiments including 
the 14 TeV LHC and ILC~\cite{Baer:2013cma}.}  
\label{mB-BR}
\end{center} 
 \end{figure}

Since $M_{X_s} < M_\phi/2$ as placed by the LUX2016 limit 
in Eq.(\ref{mB:limit:LUX}), the DSSM can be constrained by 
the Higgs invisible decay searched at collider experiments.  
As the Higgs $\phi$ acts as a pseudo dilaton,
the Higgs-onshell coupling to $X_s \bar{X}_s$, relevant to 
the invisible decay process, should be the same as 
that determined by the low-energy theorem, i.e., $q \sim M_\phi \ll v$ 
in Eq.(\ref{3-point}). 
The partial decay width of the Higgs $\phi$ to the $X_s \bar{X}_s$ 
is thus unambiguously computed from Eq.(\ref{3-point}) to be 
\begin{equation}
 \Gamma(\phi \to X_s \bar{X}_s) 
= \frac{\lambda_{X_s}^2 v^2}{4 \pi M_\phi} 
\sqrt{1 - \frac{4 M_{X_s}^2}{M_\phi^2}}
\,. 
\end{equation} 
The branching ratio is then constructed as 
$
{\rm Br}[\phi \to X_s \bar{X}_s] 
=
\Gamma(\phi \to X_s \bar{X}_s)/\Gamma_\phi^{\rm tot}  
= 
\Gamma(\phi \to X_s\bar{X}_s)/
[\Gamma_\phi^{\rm SM} + \Gamma(\phi \to X_s \bar{X}_s)]
$, 
with the total SM Higgs width (without the $X_s \bar{X}_s$ decay mode) 
$\Gamma_\phi^{\rm SM}\simeq 4.1$ MeV at the mass of 125 GeV~\cite{tot:higgs}. 
The currently most stringent upper limit on the Higgs invisible decay 
has been set by the CMS Collaboration combined with 
the run II data set with the luminosity of 
$2.3\,{\rm fb}^{-1}$~\cite{CMS:2016rfr}. 
Figure~\ref{mB-BR} shows the exclusion limit on the $X_s$ mass 
at 95\% C.L., 
 ${\rm Br}_{\rm invisible} \lesssim 0.2$~\cite{CMS:2016rfr}. 
 From the figure, one reads off the upper limit, 
\begin{equation} 
 M_{X_s} \lesssim 18 \, {\rm GeV}
\,,\label{mB:limit:BR}
\end{equation} 
which is milder than the constraint 
from the direct detection experiment in Eq.(\ref{mB:limit:LUX}).  
In Fig.~\ref{mB-BR} the future prospected 95\% C.L. limits 
in the LHC and ILC experiments~\cite{Baer:2013cma} have also been shown.  
We thus see that the 14 TeV LHC with the luminosity of 300 ${\rm fb}^{-1}$ 
has the potential to exclude, or detect 
the DSSM with the mass up to $\simeq 14$ GeV, 
which is slightly less sensitive compared to the direct detection experiments.  
The ILC with higher statistics will be more sensitive to reach the exclusion 
and discovery sensitivity for the mass up to $\simeq 6$ GeV.

\section{DSSM in thermal history} 
\label{thermal}

The SM Higgs Lagrangian with the HLS in Eq.(\ref{Lag}) is formulated in the 
vacuum where the electroweak symmetry is broken,  
hence in the thermal history of the universe 
the DSSM emerges after the electroweak phase transition at the temperature 
$T = {\cal O}(v)$. 
At that time the DSSM was in the thermal (chemical) equilibrium with 
the photon and other SM particles due to the Higgs portal coupling in Eq.(\ref{3-point}).  
As the universe cooled down to be at the temperature $x=M_{X_s}/T \sim 1$, 
the DSSM $X_s$ became nonrelativistic and 
the number density begins to fall down like $\sim e^{-M_{X_s}/T}$.   
Then the DSSM density gets so diluted due to the expansion of the universe, 
to make the DSSM cease to interact, and  
finally freezes out at $x_f=M_{X_s}/T_f (
=  {\cal O}(10))$. 
Below $T=T_f$ ($x>x_f$) the DSSM number density per comoving volume 
stays to be constant and the DSSM cools down  
to become a cold dark matter just like WIMPs,  
with the relic abundance observed in the universe today.

Such a relic abundance can be estimated by the standard 
procedure, so-called the freeze-out thermal relic~\cite{KT}: 
$ \Omega_{X_s} h^2 
= 
\frac{2 \times (1.07 \times 10^9)x_f}
{g_*(T_f)^{1/2} M_{\rm Pl} {\rm GeV} \langle 
\sigma_{\rm ann} v_{\rm rel} \rangle} 
$, 
where $M_{\rm pl}$ stands for the Planck mass scale $\simeq 10^{19}$ GeV, 
$\langle \sigma_{\rm ann} v_{\rm rel} \rangle $ 
is the thermal average of 
the annihilation cross section  
times the relative velocity of $X_s \bar{X}_s$, 
$v_{\rm rel}$,  
and $g_*(T_f)$ denotes the effective degrees of freedom for relativistic particles at $T=T_f$. 
The prefactor 2 comes from counting both $X_s$-particle 
and $\bar{X}_s$-anti-particle
 present today. 
The freeze out temperature $T_f$ can be determined by 
$ 
x_f = \ln [2 \times 0.038 \times [g_*(T_f) x_f]^{-1/2} M_{\rm pl}\cdot M_{X_s} \cdot 
\langle \sigma_{\rm ann} v_{\rm rel} \rangle  ] 
$.

Since the freeze-out temperature 
$(T_f \sim M_{X_s}/10 
={\cal O}(1)\,{\rm GeV})$ 
is expected to be much smaller than the Higgs mass scale, 
the Higgs portal coupling formula in Eq.(\ref{3-point}), 
derived from the low-energy theorem for the scale  symmetry, 
is applicable to estimate the freeze-out of the Higgs portal process. 
The explicit computation actually shows that 
the Higgs portal process decouples from the thermal equilibrium at 
$x_f |_\textrm{Higgs portal} \simeq 10$ for $M_{X_s}={\cal O}(10)$GeV.

In addition to the Higgs portal process, 
one should note that 
the DSSM is essentially a soliton, an 
extended particle with 
a finite radius.  
The $X_s \bar{X_s}$ 
``annihilation'' into the $U(1)_{X_s}$ current  
can be viewed as the classical $X_s \bar{X}_s$ collision 
with the $U(1)_{X_s}$ charge radius    
$\langle r_{_{X_s}}^2 \rangle_{X_s}$ and 
the black disc approximation could be applied, so   
\begin{equation} 
\langle \sigma_{\rm ann} v_{\rm rel} \rangle_{\rm radius} 
\simeq 
4 \pi \cdot \langle r_{_{X_s}}^2 \rangle_{X_s} 
\,, 
\end{equation}   
(Similar observation was made 
in Ref.~\cite{Kitano:2016ooc}.) 
As it will turn out, 
the freeze out of this classical collision is actually later than 
the Higgs portal process, 
$x_f |_{\rm radius} \simeq 20 > x_f |_\textrm{Higgs portal}$ 
for $M_{X_s}={\cal O}(10)$ GeV, 
so the relic abundance of the DSSM is determined by this classical collision~\footnote{
 In the sense of magnitude relation on the reaction rates,  
the present DSSM-production scenario 
therefore looks similar to the self-interacting dark matter scenario 
in which the dark-sector self-interaction gets dominant 
for the thermal-freeze out relic.}.

\begin{figure}[t]
\begin{center}
   \includegraphics[scale=0.4]{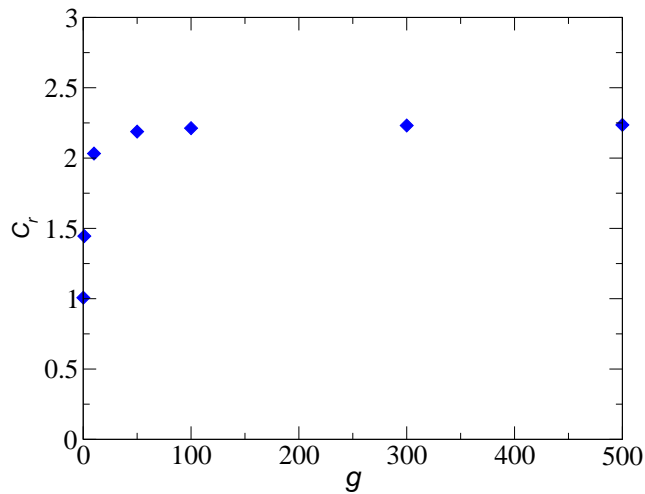} 
\caption{ 
Numerical result for the coefficient of the topological charge radius $C_r$ as a function of $g$.  }   
\label{radius}
\end{center} 
 \end{figure}

To estimate the size of $\langle r^2_{_{X_s}} \rangle_{X_s}$,  
{\it as a simple benchmark}, we shall take the infinite SM$\rho$ mass limit, $a \rightarrow \infty$ with $g=$ fixed, 
Eq.(\ref{solitoneq2}). 
Then the skyrmion system is reduced to essentially the same as the one  
analyzed in Ref.~\cite{Kitano:2016ooc} 
(up to the scale-non-invariant form of the Higgs profile,  
which we have checked does not affect the soliton 
solution and the mean radius). 
The soliton mass $M_{X_s}$ is similar to that already given in Ref.~\cite{Kitano:2016ooc}, while
 the square of the mean radius $\langle r_{_{X_s}}^2 \rangle_{X_s}$ 
 (not shown in Ref.~\cite{Kitano:2016ooc}) is calculated by the standard
 formula~\cite{Jackson:1983bi, Adkins:1983ya}:
 \begin{eqnarray}
 \langle r_{_{X_s}}^2 \rangle_{X_s} &=& \int_0^\infty r^2dr \rho_{X_s}(r) 
 \equiv \left( \frac{C_r}{gv} \right)^2, \notag \\
\rho_{X_s}(r) &=& 4\pi r^2 X_s^0(r), 
  \end{eqnarray}
where $X_s^\mu$ is the topological charge current 
$X_s^\mu=\frac{\epsilon^{\mu\nu\alpha\beta}}{24\pi^2} 
{\rm Tr}[(U^\dagger\partial_\nu U)(U^\dagger\partial_\alpha U)(U^\dagger\partial_\beta U)]$.
Our result for $C_r$ is given in Fig.\ref{radius}.

 Thus in the Skyrme term limit $a\rightarrow \infty$ with $g=$ fixed  as our benchmark case,  we have  
\begin{eqnarray}
\langle r_{_{X_s}}^2 \rangle_{X_s} 
&\simeq& 
\frac{(2.2)^2 
}{g^2 v^2} 
\,, \qquad
M_{X_s} \simeq \frac{35v}{g} \simeq 17 \, {\rm GeV}\times \left( \frac{500}{g}  \right)\,, \nonumber \\ 
&& 
\Omega_{X_s} h^2 \simeq 0.1 \times \left( \frac{500}{g}  \right)^{-2}
\,, \label{r-m} 
\end{eqnarray} 
where the estimate of the thermal relic abundance 
accumulated by the black disc collision was made, and 
the freeze out temperature $x_f=M_{X_s}/T_f \simeq 20$ is computed to be
 almost constant in the mass and the radius, and we have used 
 $g_*(T_f)=100$ as the effective degree of freedom at the freeze out
. 
The benchmark value 
$g = 500\, (\gg 1)$ corresponds to 17 GeV 
as a reference value of  the upper bound from the direct and 
indirect searches in the region $M_{X_s} = {\cal O}\, (10)$  GeV.  
Although we have two parameters $(a,g)$ for the  generic HLS case Eq.(\ref{soliton}), Eq.(\ref{r-m}) has only one parameter $g$ in our benchmark case Eq.(\ref{solitoneq2}) in the Skyrme term limit $a\rightarrow \infty$ with $g=$ fixed.
Note that the value $g\sim
500$ from 
$M_{X_s} = 
{\cal O} (10)$  GeV in our 
benchmark case  trivially satisfies 
the constraint from the vector boson scattering, 
$g \gtrsim 0.4$~\cite{Kitano:2016ooc}. 
%
%
%
%
%
%
Thus we find 
$ 
\Omega_{X_s} h^2\, = {\cal O}(0.1) 
$
for 
$ 
M_{X_s} = {\cal O}(10)\, {\rm GeV}
$, which is roughly consistent with 
the presently observed 
dark matter relic $\simeq 0.12$~\cite{Ade:2013zuv}.

Since the DSSM has a conserved skyrmion number,
there may be an alternative way to asymmetrically generate the current relic abundance of the DSSM 
through the electroweak sphaleron process together with both the DSSM and the SM  
Higgs Lagrangian. This possibility will be pursued elsewhere.


\section{Conclusion and discussion} 

We have shown a novel possibility that 
the dark matter candidate  
exists already within the 
Standard Model (SM), not beyond it, 
through nonperturbative dynamics.
The SM Higgs Lagrangian 
is  cast into precisely the scale-invariant nonlinear sigma model, with the SM Higgs being 
a  pseudo dilaton. 
It is further shown to be
gauge equivalent to the scale-invariant version of the hidden local symmetry (HLS) Lagrangian whose dynamical gauge boson 
``SM-rho meson (SM$\rho$)'' stabilizes the skyrmion,
 ``Dormant Skyrmion in the SM (DSSM)'' $X_s$, 
 a novel candidate for the dark matter 
 without explicit recourse to possible underlying theory.

The scale invariance of the whole dynamics is essential, 
which unambiguously determines the couplings of the DSSM to the SM Higgs as a pseudo-dilaton 
in terms of the low energy theorem of the
spontaneously broken scale invariance. 
This imposes a definite constraint on the mass $M_{X_s} \lesssim 13$ GeV 
from the direct detection experiments~\cite{Akerib:2016vxi}. 
With such a mass smaller than half of the SM Higgs mass,  
we have also found the constraint 
$M_{X_s} \lesssim 18$ GeV from the SM Higgs 
invisible decay data~\cite{CMS:2016rfr} definitely 
by  the low-energy theorem.

Based on this salient constraint we have discussed the thermal history of the 
DSSM, and estimated the thermal relic abundance, in view of the extended size of the soliton, 
in some  benchmark cases including the heavy SM$\rho$ mass limit (scale-invariant Skyrme model limit).  
It was shown that the estimated present relic density  of the DSSM 
with the mass of ${\cal O}(10\,{\rm GeV})$ 
is roughly consistent with the observed-cold dark matter relic-abundance 
$\Omega_{\rm cdm} h^2 \simeq 0.12$.

Although our rough estimate on the relic abundance was made only in the benchmark cases, 
including the heavy SM$\rho$ mass limit $a\rightarrow \infty$ with $g=$ fixed, 
more interesting would be the lighter SM$\rho$ mass case. 
There have been many studies on this case in the hadron physics~\cite{Igarashi:1985et,Park:2003sd}. 
In the forthcoming paper 
we will report explicit calculations of
the size and mass of the DSSM for the whole parameter space $(g,a)$.

We have also discussed discovery/exclusion possibilities in the future experiments.  
Both the prospected direct detection and indirect detection experiments 
have the sensitivity enough to discover the DSSM with the mass of ${\cal O}(10\,{\rm GeV})$ 
within a couple of decades (See Figs~\ref{mB-sigma} and \ref{mB-BR}).  
If the DSSM signals were not observed, 
conversely, the DSSM mass could be constrained 
by future experiments to be lower and lower, 
so that the relic abundance could get larger. 
Then the DSSM might exceed 
the present cold-dark matter density, 
which would imply necessity to go beyond the SM. 
At any rate, the future-prospected experiments will  
clarify whether or not the DSSM can 
explain the dark matter presently observed in our universe.

\acknowledgments 
We are grateful to Masayasu Harada for useful comments on 
the skyrmion. 
We also thank Andreas Karch, Taichiro Kugo,
and Tadakatsu Sakai for valuable discussions. 
This work was supported in part by 
the JSPS Grant-in-Aid for Young Scientists (B) \#15K17645 (S.M.), 
and the work of H.O. is supported by the RIKEN Special Postdoctoral
Researcher program.


\begin{thebibliography}{99} 




\bibitem{Yamawaki:2016kdz} 
  K.~Yamawaki,
  PTEP {\bf 2016}, no. 6, 06A107 (2016).




\bibitem{Bando:1987br} 
  M.~Bando, T.~Kugo and K.~Yamawaki,
  Phys.\ Rept.\  {\bf 164}, 217 (1988).





\bibitem{Fukano:2015zua} 
  H.~S.~Fukano, S.~Matsuzaki, K.~Terashi and K.~Yamawaki,
  Nucl.\ Phys.\ B {\bf 904}, 400 (2016).
  See also, 
  K.~Yamawaki,
  arXiv:1511.06883 [hep-ph]; 
  PTEP {\bf 2016}, no. 6, 06A107 (2016).
  
  
  
   
\bibitem{Bando:1984ej} 
 M.~Bando, T.~Kugo, S.~Uehara, K.~Yamawaki, and T.~Yanagida, 
Phys. Rev. Lett. 
 {\bf 54} (1985) 1215;
M.~Bando, T.~Kugo, and K.~Yamawaki, 
Nucl. Phys. 
 {\bf B259}  (1985) 493.

\bibitem{Harada:2003jx} 
  M.~Harada and K.~Yamawaki,
  Phys.\ Rept.\  {\bf 381}, 1 (2003).





\bibitem{Eichenherr:1978qa} 
  H.~Eichenherr,
  Nucl.\ Phys.\ B {\bf 146}, 215 (1978)
  Erratum: [Nucl.\ Phys.\ B {\bf 155}, 544 (1979)];
%
  V.~L.~Golo and A.~M.~Perelomov,
  Phys.\ Lett.\  {\bf 79B}, 112 (1978);
%
  A.~D'Adda, M.~Luscher and P.~Di Vecchia,
  Nucl.\ Phys.\ B {\bf 146}, 63 (1978);
  Nucl.\ Phys.\ B {\bf 152}, 125 (1979);
%
  E.~Witten,
  Nucl.\ Phys.\ B {\bf 149}, 285 (1979);
%
  I.~Y.~Arefeva and S.~I.~Azakov,
  Nucl.\ Phys.\ B {\bf 162}, 298 (1980).
%
  H.~E.~Haber, I.~Hinchliffe and E.~Rabinovici,
  Nucl.\ Phys.\ B {\bf 172}, 458 (1980).

\bibitem{Weinberg:1997rv} 
  S.~Weinberg,
  Phys.\ Rev.\ D {\bf 56}, 2303 (1997).
  
  
  



  
\bibitem{Igarashi:1985et} 
  Y.~Igarashi, M.~Johmura, A.~Kobayashi, H.~Otsu, T.~Sato and S.~Sawada,
  Nucl.\ Phys.\ B {\bf 259}, 721 (1985);
  U.~G.~Meissner and I.~Zahed,
  Phys.\ Rev.\ Lett.\  {\bf 56}, 1035 (1986);
  U.~G.~Meissner, N.~Kaiser, A.~Wirzba and W.~Weise,
  Phys.\ Rev.\ Lett.\  {\bf 57}, 1676 (1986);
  F.~R.~Klinkhamer,
  Z.\ Phys.\ C {\bf 31}, 623 (1986).
  %
  
  \bibitem{Park:2003sd} 
  B.~Y.~Park, M.~Rho and V.~Vento,
  Nucl.\ Phys.\ A {\bf 736}, 129 (2004);
  Nucl.\ Phys.\ A {\bf 807}, 28 (2008).
  Y.~L.~Ma, M.~Harada, H.~K.~Lee, Y.~Oh, B.~Y.~Park and M.~Rho,
  Phys.\ Rev.\ D {\bf 90}, no. 3, 034015 (2014);
  B.~R.~He, Y.~L.~Ma and M.~Harada,
  Phys.\ Rev.\ D {\bf 92}, no. 7, 076007 (2015).

\bibitem{Kitano:2016ooc} 
  R.~Kitano and M.~Kurachi,
  JHEP {\bf 1607}, 037 (2016).
  
\bibitem{Carruthers:1971vz} 
 See e.g.,  P.~Carruthers,
  Phys.\ Rept.\  {\bf 1}, 1 (1971).


 
\bibitem{Akerib:2015rjg} 
  D.~S.~Akerib {\it et al.} [LUX Collaboration],
  Phys.\ Rev.\ Lett.\  {\bf 116}, no. 16, 161301 (2016).

\bibitem{Akerib:2016vxi} 
  D.~S.~Akerib {\it et al.},
   arXiv:1608.07648  [astro-ph.
   CO]; 
  A.~Tan {\it et al.}    [PandaX-II Collaboration],
   Phys.\ Rev.\ Lett.\  {\bf 117}, no. 12, 121303 (2016).







\bibitem{Harvey:1996ur} 
See e.g.,  
J.~A.~Harvey,
  In *Trieste 1995, High energy physics and cosmology* 66-125
  [hep-th/9603086].

    


\bibitem{Kurachi:2014qma} 
  M.~Kurachi, S.~Matsuzaki and K.~Yamawaki,
  Phys.\ Rev.\ D {\bf 90}, no. 5, 055028 (2014).

   
 

\bibitem{Eguchi:1976iz} 
  T.~Eguchi,
  Phys.\ Rev.\ D {\bf 14}, 2755 (1976);
  T.~Kugo,
  Prog.\ Theor.\ Phys.\  {\bf 55}, 2032 (1976). This method was fully developed  by 
  W.~A.~Bardeen, C.~T.~Hill and M.~Lindner,
  Phys.\ Rev.\ D {\bf 41}, 1647 (1990) 
  for  the top quark condensate model proposed by   
  V.~A.~Miransky, M.~Tanabashi and K.~Yamawaki,
  Phys.\ Lett.\ B {\bf 221}, 177 (1989);
  Mod.\ Phys.\ Lett.\ A {\bf 4}, 1043 (1989) 
 based on the explicit  (gauged) Nambu-Jona-Lasinio model.
  



\bibitem{Alonso:2016oah} 
See, e.g.,  R.~Alonso, E.~E.~Jenkins and A.~V.~Manohar,
  JHEP {\bf 1608}, 101 (2016), and references therein.


\bibitem{Skyrme:1961vq} 
  T.~H.~R.~Skyrme,
  Proc.\ Roy.\ Soc.\ Lond.\ A {\bf 260}, 127 (1961).




 
  



\bibitem{Ohki:2008ff} 
  H.~Ohki {\it et al.} [JLQCD Collaboration],
  Phys.\ Rev.\ D {\bf 78}, 054502 (2008);
  Phys.\ Rev.\ D {\bf 87}, 034509 (2013).
 
  
\bibitem{Hisano:2015rsa} 
  J.~Hisano, K.~Ishiwata and N.~Nagata,
  JHEP {\bf 1506}, 097 (2015).
  
  
\bibitem{Feng:2014uja} 
  J.~L.~Feng {\it et al.},
  arXiv:1401.6085 [hep-ex].

\bibitem{Billard:2013qya} 
  J.~Billard, L.~Strigari and E.~Figueroa-Feliciano,
  Phys.\ Rev.\ D {\bf 89}, no. 2, 023524 (2014).




\bibitem{tot:higgs}
https://twiki.cern.ch/twiki/bin/view/
LHCPhysics/CERNYellowReportPageBR
\#TotalWidthAnchor.



\bibitem{CMS:2016rfr} 
  CMS Collaboration [CMS Collaboration],
  CMS-PAS-HIG-16-016.



\bibitem{Baer:2013cma} 
  H.~Baer {\it et al.},
  arXiv:1306.6352 [hep-ph].


\bibitem{KT} 
E.W. Kolb, M.S. Turner, The Early Universe, Addison-Wesley, 1990. 


\bibitem{Jackson:1983bi} 
  A.~D.~Jackson and M.~Rho,
  Phys.\ Rev.\ Lett.\  {\bf 51}, 751 (1983).
  doi:10.1103/PhysRevLett.51.751

\bibitem{Adkins:1983ya} 
  G.~S.~Adkins, C.~R.~Nappi and E.~Witten,
  Nucl.\ Phys.\ B {\bf 228}, 552 (1983).
  doi:10.1016/0550-3213(83)90559-X


\bibitem{Ade:2013zuv} 
  P.~A.~R.~Ade {\it et al.} [Planck Collaboration],
  Astron.\ Astrophys.\  {\bf 571}, A16 (2014).






  

\end{thebibliography}
\end{document}